\pdfoutput=1
\documentclass[conference]{IEEEtran}
\usepackage{cite}
\usepackage{amsmath,amssymb,amsfonts}
\usepackage{graphicx}

\begin{document}

\bstctlcite{IEEEexample:BSTcontrol}
\title{Sub-100KB On-Device CNN Training via Dynamic Sparse Updates and Hardware--Software Co-Design
}

\author{\IEEEauthorblockN{I-Hsuan Li and Tian-Sheuan Chang}\\
\IEEEauthorblockA{\textit{Institute of Electronics, National Yang Ming Chiao Tung University,}
Hsinchu, Taiwan\\
}
\IEEEauthorblockA{Email: \{shelly.ee11, tschang\}@nycu.edu.tw}
}

\maketitle

\begin{abstract}
The increasing demand for real-time processing and personalized deep learning has made edge-device training essential but challenging due to high memory and computational costs. To address this, we propose a Convolutional Neural Network (CNN) training accelerator with a dynamic sparse update mechanism that selectively updates important parameters using a random channel strategy, avoiding costly evaluations. Combined with activation pruning, our approach reduces activation storage to just 3.4\% of the original. 
Our hardware features a fused forward dataflow that concurrently processes group normalization and convolution, reducing training time by 45\% through computation reordering and efficient zero-skipping. Implemented in TSMC 28nm at FP8 precision, the design operates at 800 MHz, achieving 8 TOPS peak throughput, 31.13 TOPS/W energy efficiency, and 14.8 TOPS/mm$^2$ area efficiency, with power consumption ranging from 257 to 272 mW. The design enables on-device training within a sub-100KB on-chip SRAM budget (95KB in silicon).

\end{abstract}

\section{Introduction}
\label{sec:introduction}

On-device training is important for privacy-preserving and personalized edge AI, but it is much more memory-intensive than inference because backpropagation must store intermediate features and gradients. This is challenging for edge chips with only hundreds of kilobytes to a few megabytes of SRAM (often around 256\,KB). For example, training MobileNetV2 (3.5M parameters) requires 75.65\,MB for a single batch, forcing frequent DRAM access and high power consumption. Therefore, memory-efficient training algorithms and hardware co-design are essential for practical on-device learning.

For memory-efficient training, model pruning is widely used, including structured~\cite{li2016pruning, he2017channel, fang2023depgraph}, unstructured~\cite{gale2019state}, and sub-structured pruning~\cite{wang2022swpu, niu2020patdnn}. Structured methods are hardware-friendly but usually provide lower compression, while irregular sparsity complicates efficient implementation. Another direction updates only selected parameters, such as the classifier~\cite{sharif2014cnn}, biases~\cite{cai2020tinytl}, or normalization layers~\cite{frankle2020training}, but this limited adaptation scope often reduces accuracy. Methods such as SparseUpdate~\cite{lin2022device} and TinyTrain~\cite{kwon2023tinytrain} improve update selection; however, SparseUpdate depends on prior target-dataset knowledge and TinyTrain requires online full backpropagation and activation storage, resulting in high memory and power overhead.

Hardware training accelerators also exploit sparsity through dynamic pruning or sparse execution, as in SWPU~\cite{wang2022swpu}, Trainer~\cite{wang2022trainer}, TSUNAMI~\cite{kim2022tsunami}, and GANPU~\cite{kang20207}. However, many of these designs require complex sparsity control and prediction logic, which increases implementation cost; for example, SWPU reports up to 50\% power in sparse-processing hardware. HNPU~\cite{han2023hnpu} supports multi-precision and reconfiguration, but its area utilization and power efficiency remain limited for tight edge constraints.

To address these challenges, we propose a CNN training accelerator with dynamic sparse updates. Our contributions are threefold:
\begin{enumerate}
    \item A dynamic sparse update algorithm that uses random channel selection to avoid complex prediction logic while retaining effective adaptation.
    \item A fused forward dataflow that overlaps group normalization (GN) and convolution in early frozen layers to reduce redundant operations.
    \item A row-wise processing scheme that enables low-overhead zero-skipping for both weights and activations.
\end{enumerate}
The resulting design supports on-device training within a 100\,KB SRAM buffer, with high area efficiency and competitive energy efficiency.

The remainder of this paper is organized as follows. Section~II presents the model optimization and update scheme. Section~III describes the hardware architecture and dataflow. Section~IV reports experimental results, and Section~V concludes this work.

\section{On-Device Training Methodology}
\label{sec:methodology}

\begin{figure}[htbp]
\centering
\includegraphics[height=!,width=1\linewidth,keepaspectratio=true]{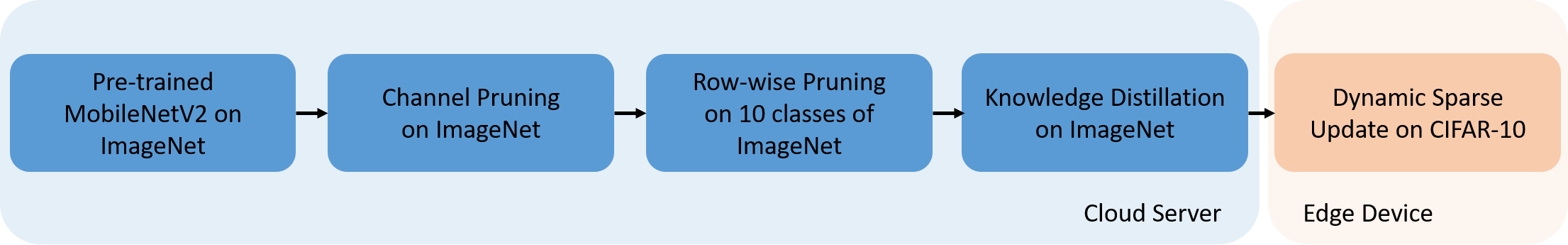}
\caption{The overall flow-chart with the network pruning and dynamic sparse update procedure \cite{li2024dynamic} with the MobileNetv2 as base model to train the downstream task, CIFAR-10.}
\label{03_flowchart}
\end{figure}

Effective on-device training accelerators require models that are lightweight, memory-efficient, and hardware-friendly; therefore, as shown in Fig.~\ref{03_flowchart}, we adopt a hybrid flow that combines offline and online optimization for a pretrained MobileNetV2. In the offline stage, mixed-level pruning builds a hardware-friendly sparse backbone, reduces model size by 95.2\%, and identifies critical update regions, followed by knowledge distillation to recover accuracy on the source dataset (e.g., ImageNet). In the online stage, the accelerator applies dynamic sparse updates with random channel selection in the selected layers to minimize memory/computation overhead, and the resulting compact model is fine-tuned on the target task. The following subsections detail each component.

\subsection{Hardware-Aware Architectural Optimization}

This subsection presents the offline, hardware-aware optimizations used to reduce model size, memory traffic, and computation cost through BN-to-GN replacement, GN-Conv reordering, and mixed-level pruning.

\subsubsection{GN-Conv Reordering and fused forward computation for frozen layers}
MobileNetV2 originally uses BN, but BN is unsuitable for SRAM-limited on-device training because it depends on reliable mini-batch statistics; even progressive BN approaches~\cite{PBN} are still designed for mini-batch regimes. For our batch-size-1 setting, we therefore adopt GN, which is batch-size agnostic and more stable under small batches~\cite{wu2018group}. In addition, conventional post-GN ordering (Conv$\rightarrow$GN) is hardware-inefficient because GN must wait for full convolution outputs before statistics are available, creating a latency bottleneck.

\begin{figure}[htbp]
\centering
\includegraphics[height=!,width=0.7\linewidth,keepaspectratio=true]{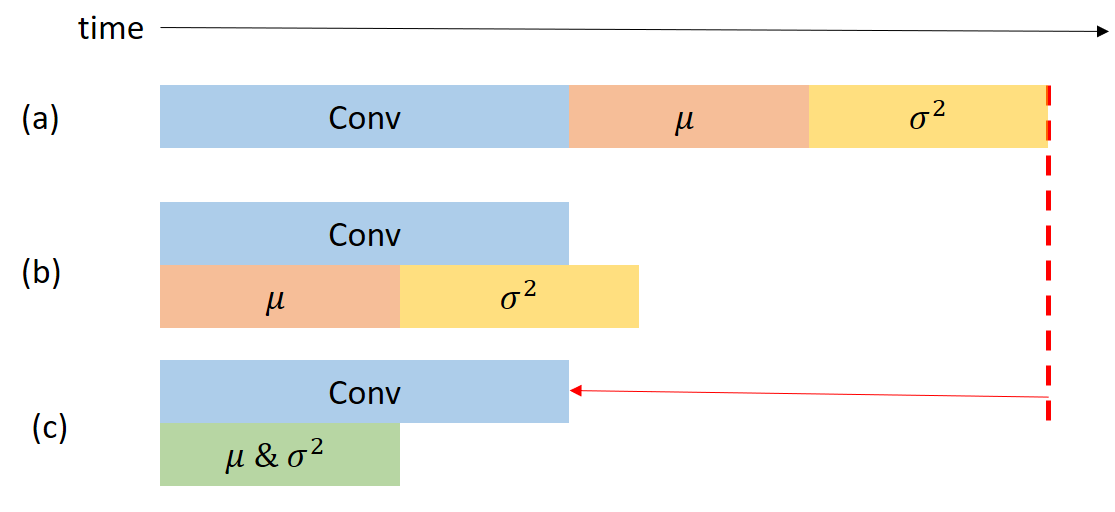}
\caption{(a) original calculation method, (b) fused forward calculation method with GN-Conv reordering, and (c) enhanced fused forward method with simultaneous mean and variance calculation}
\label{03_gntime}
\end{figure}

To mitigate this bottleneck, we reorder the pipeline to pre-GN (GN$\rightarrow$Conv) and implement a fused forward dataflow so that GN statistics are accumulated concurrently with convolution, reducing end-to-end latency as shown in Fig.~\ref{03_gntime}. The fused formulation is:

\begin{align}
y_{i,j,k} &= \sum_{m,n,c} w_{m,n,c,k} \cdot \text{GN}(x_{i',j',c}) \nonumber \\
&= \sum_{m,n,c} w_{m,n,c,k} \cdot \left(\gamma \cdot \frac{x_{i',j',c} - \mu_{g}}{\sqrt{\sigma_{g}^{2}+\epsilon}} + \beta\right) \nonumber \\
&= \frac{\gamma}{\sqrt{\sigma_{g}^{2}+\epsilon}} \left( \text{Conv}(x, w) - \mu_g \cdot W_{\text{sum}} \right) + \beta \cdot W_{\text{sum}}
\label{eq:gn_conv_fusion}
\end{align}
where $y_{i,j,k}$ is the output, $x_{i',j',c}$ is the corresponding input patch, and $W_{\text{sum}} = \sum_{m,n,c} w_{m,n,c,k}$.

For frozen layers, $W_{\text{sum}}$ is precomputable, so GN-related terms are folded into lightweight affine corrections and normalization overhead is largely hidden inside convolution. In addition, variance evaluation is parallelized by concurrently accumulating $\sum x_i$ and $\sum x_i^2$, then finishing $\sigma_g^2$ with a single subtraction after $\mu_g$ is ready. This fused-and-parallel schedule shortens the critical path and yields the latency reduction shown in Fig.~\ref{03_gntime}.

\subsubsection{Mixed-Level Model Pruning}

To create a compact base model, we adopt a three-stage offline pruning flow: layer-wise, channel-wise, and row-wise pruning (Table~\ref{tab:pruning_results}). We first remove the final bottleneck stage of MobileNetV2, reducing convolutional weights from 2.2M to 1.7M with negligible transfer-learning impact. We then apply DepGraph channel pruning~\cite{fang2023depgraph}, further reducing the weight count to 0.7M (68\% cumulative reduction).

\begin{figure}[!t]
\centering
\includegraphics[width=0.9\columnwidth]{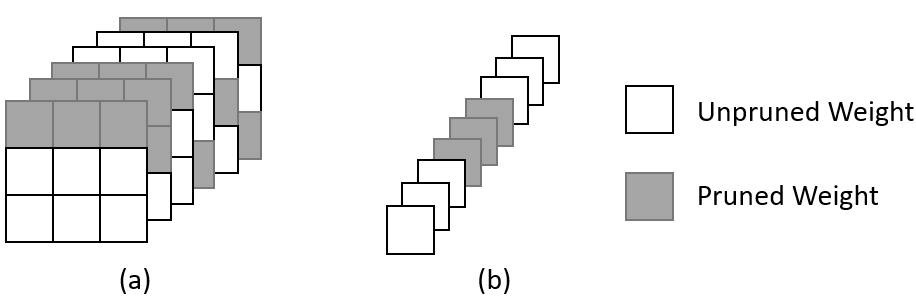}
\caption{Row-wise weight pruning for (a) $3 \times 3$ and (b) $1 \times 1$ filters.}
\label{fig:row_pruning}
\end{figure}

In the final stage, row-wise pruning removes full rows in each filter (Fig.~\ref{fig:row_pruning}), which directly matches row-wise zero-skipping hardware. For $1 \times 1$ filters, rows are defined along grouped channel dimensions. We iteratively prune 10\% and retrain until saturation, while protecting sensitive layers (including the first and all depth-wise layers~\cite{evci2020rigging}), yielding a final model size of about 0.14M weights. To further improve efficiency, we also apply row-wise activation pruning, so the same lightweight zero-skipping logic can be shared by both weights and activations.

\subsection{Dynamic Sparse Update}
\label{sec:dynamic_update}

Even after offline pruning, storing activations and gradients for backpropagation remains the dominant memory bottleneck. To meet strict on-chip constraints, we use a dynamic sparse update strategy with three components: memory-aware layer selection, dynamic channel selection, and a three-phase schedule for stable convergence.

\subsubsection{Layer Selection for Memory-Constrained Updates}
Under a fixed memory budget, updating later layers gives higher accuracy because these layers are more task-specific, while early layers are more transferable and can remain frozen. Therefore, we allocate memory from the last layer backward and keep adding layers until the total storage for weights, activations, and gradients reaches the target budget.

\subsubsection{Dynamic Channel Selection}
Within selected layers, we update a random channel subset each epoch using update ratio $r$, instead of fixing channels throughout training. This channel rotation increases parameter coverage over time, improves final accuracy, and is inexpensive to implement in hardware.

\subsubsection{Three-Phase Training Schedule}
To balance adaptation and convergence, we use three training phases:
\begin{itemize}
    \item \textbf{Initial (Fixed):} update a fixed channel set to establish a stable adaptation baseline.
    \item \textbf{Middle (Dynamic):} enable random channel rotation for broader exploration.
    \item \textbf{Final (Fixed):} switch back to fixed channels for stable fine-tuning.
\end{itemize}
This fixed-dynamic-fixed schedule improves exploration in mid-training while retaining stable convergence, and allows broader parameter updates than static masks under the same memory budget.

\section{Hardware Implementation}
\label{sec:hardware}

\subsection{Architectural Overview and Dataflow}

Based on the proposed model optimization and dynamic sparse update strategy, we design a low-power training accelerator with 24 row-wise PE blocks, 24 accumulators, dedicated GN modules, and four APUs for ReLU6, activation pruning, and sparse data encoding. The design uses FP8 for both compute and storage, and adopts row-wise sparse scheduling (following~\cite{chang2019vwa}) for low-overhead zero-skipping.

\begin{figure}[!t]
\centering
\includegraphics[width=\columnwidth]{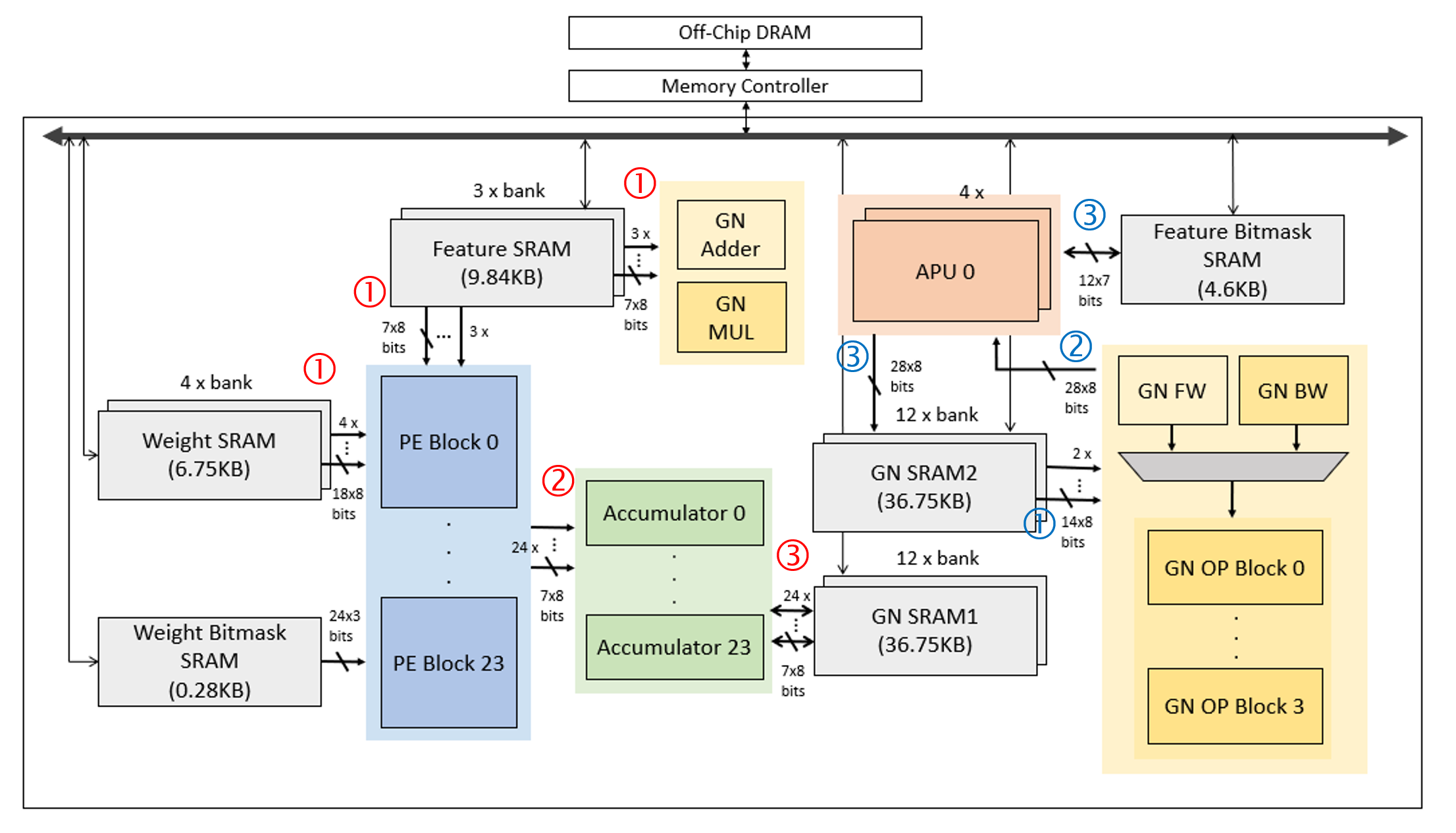}
\caption{Top-level dataflow used in the accelerator, including the two-stage fused forward flow for frozen layers.}
\label{fig:fused_flow}
\end{figure}

The accelerator uses mixed dataflow: fused forward flow for frozen early layers, and general forward/backward flow for updated layers. In fused mode (Fig.~\ref{fig:fused_flow}), Stage 1 performs convolution while GN1 accumulates mean/variance statistics, and Stage 2 completes GN2 and APU operations. Two multi-bank GN SRAMs operate as ping-pong buffers, with bitmask SRAMs for both weights and features. For updated layers, the forward path follows GN $\rightarrow$ Conv $\rightarrow$ ReLU6, while backward propagation reverses the sequence as ReLU6 gradient $\rightarrow$ Conv gradient $\rightarrow$ GN gradient.

\subsection{GN Hardware}
The GN hardware is divided into GN1 (statistic accumulation) and GN2 (final normalization/gradient operations).

\subsubsection{GN1 Module}
GN1 contains a \textbf{GN Adder} and a \textbf{GN MUL}. The GN Adder uses a 9-adder tree to sum 21 inputs in 3 cycles, and GN MUL has 7 multipliers. Together they generate the accumulation terms needed by GN in both passes.
\begin{itemize}
    \item \textbf{Forward:} GN MUL computes $x_i^2$ while GN Adder computes $\sum x_i$, then accumulates $\sum x_i^2$.
    \item \textbf{Backward:} GN MUL computes $\frac{\partial L}{\partial y_i} \cdot \hat{x}_i$ while GN Adder computes $\sum \frac{\partial L}{\partial y_i}$, then accumulates $\sum (\frac{\partial L}{\partial y_i} \cdot \hat{x}_i)$.
\end{itemize}
This design accumulates 21 values for both terms in 6 cycles.

\subsubsection{GN2 Module}
\begin{figure}[!t]
\centering
\includegraphics[width=\columnwidth]{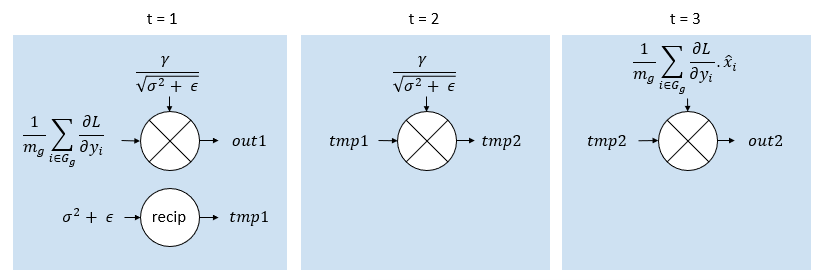}
\caption{The architecture and three-cycle calculation process of the GN BW (Backward) module, which computes the required gradients for backpropagation.}
\label{fig:gn_bw}
\end{figure}

GN2 has three units for the final GN operations.
\begin{itemize}
    \item \textbf{GN FW:} computes $\frac{\gamma}{\sqrt{\sigma_g^2 + \epsilon}}$ using an inverse square-root unit and a multiplier.
    \item \textbf{GN BW:} computes backward GN gradient terms (Fig.~\ref{fig:gn_bw}).
    \item \textbf{GN OP:} following~\cite{lu202428nm}, GN is decomposed into two MAC operations; 28 GN OP units produce 28 outputs in 2 cycles.
\end{itemize}

\begin{table*}[tbh]
\centering
\caption{Comparison with State-of-the-Art Training Accelerators (with 28\,nm-normalized area/efficiencies).}
\label{tab:sota_comparison}
\small
\setlength{\tabcolsep}{3.5pt}
\begin{tabular}{l|c|c c|c|c c|c}
\hline
\textbf{Metric} & \textbf{This work} & \multicolumn{2}{c|}{GANPU~\cite{kang20207}} & SWPU~\cite{wang2022swpu} & \multicolumn{2}{c|}{Trainer~\cite{wang2022trainer}} & HNPU~\cite{han2023hnpu} \\
\hline
Process (nm) & 28 & \multicolumn{2}{c|}{65} & 28 & \multicolumn{2}{c|}{28} & 28 \\
Max Frequency (MHz) & \textbf{800} & \multicolumn{2}{c|}{200} & 675 & \multicolumn{2}{c|}{440} & 250 \\
SRAM (KB) & \textbf{95} & \multicolumn{2}{c|}{676} & 252 & \multicolumn{2}{c|}{634} & 552 \\
Accuracy drop (\%)& 0.6(CF10) & \multicolumn{2}{c|}{NA} & $<1$ (IMGNet) & \multicolumn{2}{c|}{$<1$ (IMGNet)} & 2.09 (CF10) \\
Precision & FP8 & FP8 & FP16 & FP16 & FP8 & FP16 & SDFXP4-16 \\
Area (mm$^2$) & \textbf{0.54} & \multicolumn{2}{c|}{32.4} & 6.8 & \multicolumn{2}{c|}{20.96} & 12.96 \\
Power (mW) & 257--272 & \multicolumn{2}{c|}{647} & 16--556 & \multicolumn{2}{c|}{363} & 500--1162 \\
\hline
Peak Throughput (TOPS) & 8$^\dagger$ & 24.1 & 14.0 & 16.4 & 58.7 & 35.4 & 8.1 \\
Energy Efficiency (TOPS/W) & 31.1$^\dagger$ & 68.1 & 37.8 & 126.0 & 149.0 & 95.5 & 16.2 \\
Area Efficiency (TOPS/mm$^2$) & \textbf{14.8}$^\dagger$ & 0.74 & 0.43 & 2.4 & 2.8 & 1.69 & 0.63 \\
$^\ddagger$Area Eff.@28\,nm (TOPS/mm$^2$) & \textbf{14.8} & \textbf{4.0} & \textbf{2.3} & \textbf{2.4} & \textbf{2.8} & \textbf{1.69} & \textbf{0.63} \\
\hline
\multicolumn{8}{l}{\footnotesize $^\dagger$90\% structured weight and input sparsity. \footnotesize $^\ddagger$Area normalized with $A_{28}=A_{\mathrm{orig}}\cdot(28/\mathrm{node})^2$.} \\
\end{tabular}
\end{table*}

\begin{table}[tbh]
\centering
\caption{Comparison of Parameters and Pruning Ratios}
\label{tab:pruning_results}
\begin{tabular}{l c c}
\hline
\textbf{Method} & \textbf{Params (M)} & \textbf{Prune Ratio (\%)} \\
\hline
Baseline & 3.5 & 0 \\
+ Layer Pruning & 3.0 & 13.5 \\
+ Channel Pruning & 1.45 & 58.6 \\
+ Row-wise Pruning (Ours) & 0.17 & 95.2 \\
\hline
\end{tabular}
\end{table}

\begin{table}[tbh]
\centering
\caption{Accuracy vs. Weight Update Ratio on CIFAR-10 (100\,KB Memory)}
\label{tab:weight_ratio}
\begin{tabular}{l c c c c c}
\hline
\textbf{Weight Update Ratio} & 0.2 & 0.4 & 0.6 & 0.8 & 1.0 \\
\hline
Updatable Layers & 21 & 10 & 7 & 5 & 3 \\
Accuracy (\%) & \textbf{88.78} & 87.97 & 86.96 & 86.74 & 85.47 \\
\hline
\end{tabular}
\end{table}

\begin{table}[tbh]
\centering
\caption{Accuracy vs. Memory Constraint on CIFAR-10 (Ratio = 0.2)}
\label{tab:memory_constraint}
\scriptsize
\setlength{\tabcolsep}{3.5pt}
\begin{tabular}{l c c c c c c c c}
\hline
\textbf{Memory (KB)} & 64 & 100 & 128 & 256 & 320 & 384 & 448 & 512 \\
\hline
Updatable Layers & 13 & 21 & 27 & 39 & 43 & 45 & 47 & 49 \\
Accuracy (\%) & 87.93 & 88.78 & 89.34 & 90.01 & \textbf{90.27} & 90.23 & 89.28 & 87.55 \\
\hline
\end{tabular}
\end{table}

\begin{table}[tbh]
\centering
\caption{Comparison of Different Fine-tuning Methods on CIFAR-10}
\label{tab:dynamic_method}
\begin{tabular}{l c c}
\hline
\textbf{Method} & \textbf{Extra Memory} & \textbf{Accuracy (\%)} \\
\hline
No Fine-tuning & 0 & 9.88 \\
Update Last Layer Only & 20.12\,KB & 75.55 \\
Full Fine-tuning & 2.1\,MB & 91.60 \\
\hline
Fixed Channel Update & 100\,KB & 87.24 \\
\textbf{Dynamic Update (Ours)} & \textbf{100\,KB} & \textbf{88.78} \\
\hline
\end{tabular}
\end{table}

\begin{table}[tbh]
\centering
\caption{Accuracy (\%) Comparison on Various Downstream Datasets}
\label{tab:evaluation}
\scriptsize
\setlength{\tabcolsep}{3pt}
\begin{tabular}{l c c c c c c c c | c}
\hline
\textbf{Method} & \textbf{Cars} & \textbf{CF10} & \textbf{CF100} & \textbf{CUB} & \textbf{Flowers} & \textbf{Food} & \textbf{Pets} & \textbf{VWW} & \textbf{Avg.} \\
\hline
\cite{lin2022device} (INT8) & 52.3 & 87.0 & 63.8 & \textbf{54.7} & 85.5 & 62.2 & \textbf{77.4} & 88.4 & 71.4 \\
Ours (F32) & \textbf{60.8} & \textbf{89.4} & \textbf{68.5} & 53.3 & 88.4 & \textbf{66.6} & 76.2 & \textbf{89.6} & \textbf{74.1} \\
Ours (F8F) & 45.6 & 87.2 & 64.5 & 50.5 & \textbf{88.8} & 61.5 & 74.6 & 88.8 & 70.2 \\
\textbf{Ours (F8D)}& 58.4 & 88.8 & 67.2 & 52.1 & 87.7 & 65.1 & 75.4 & 89.4 & 73.0 \\
\hline
\end{tabular}
\end{table}

\section{Experimental Results}
\label{sec:experiments}

\subsection{Experimental Setup}
\label{sec:exp_setup}

\subsubsection{Model Preparation and Training}
We start from an ImageNet-pretrained MobileNetV2 and apply the offline flow in Section~II (layer/channel/row pruning + knowledge distillation~\cite{hao2023vanillakd}). For on-device transfer learning, we use dynamic sparse update with row-wise activation pruning (threshold 0.15), input size $224 \times 224$, 50-epoch SGD (batch size 1, zero momentum), and an initial learning rate of 0.01 with 5-epoch warm-up followed by cosine decay. Unless specified, experiments use FP8.

\subsubsection{Datasets}
To evaluate generalization, we use eight downstream datasets: Cars~\cite{krause20133d}, CIFAR-10, CIFAR-100~\cite{krizhevsky2009learning}, CUB-200-2011~\cite{wah2011caltech}, Flowers~\cite{nilsback2008automated}, Food-101~\cite{bossard2014food}, Oxford-IIIT Pet~\cite{parkhi2012cats}, and VWW~\cite{chowdhery2019visual}, consistent with prior work~\cite{lin2022device}.

\subsection{Hardware Implementation Results}
\label{sec:hw_results}
The proposed accelerator is implemented in Verilog and synthesized using TSMC 28\,nm CMOS technology. It operates at a clock frequency of 800\,MHz under a 0.9\,V supply.

The total design area is 0.54\,mm$^2$, where on-chip SRAM occupies 56\%. Total power is 257--272\,mW depending on dataflow, with SRAM as the dominant contributor (61\%), followed by the clock tree (22\%) and PE blocks (7\%).

Table~\ref{tab:sota_comparison} compares our design with state-of-the-art training accelerators. We achieve 8\,TOPS peak throughput and 31.13\,TOPS/W at 90\% structured sparsity, with area efficiency of \textbf{14.8\,TOPS/mm$^2$} in 28\,nm. Compared with prior designs, the key advantage is the balance of small die area and low SRAM footprint for edge deployment.

\subsection{Software-Level Ablation Studies}
\label{sec:sw_results}
We conduct ablation studies to validate each component of the proposed method.

\subsubsection{Model Pruning Results}
Table~\ref{tab:pruning_results} shows the effect of mixed-level pruning. Combining layer, channel, and row-wise pruning reduces MobileNetV2 from 3.5M to 0.17M parameters (95.2\%), yielding a compact base model for on-device training.

\subsubsection{Dynamic Sparse Update Memory Savings}
Dynamic sparse update substantially reduces training memory. Under a 100\,KB convolution-memory budget and 0.2 update ratio, we update the last 21 layers, reducing backpropagation activation storage from 1972\,KB (full fine-tuning) to 67\,KB (\textasciitilde$30\times$ reduction), while reducing trainable weights to 47K.

\subsubsection{Impact of Weight Update Ratio and Memory Budget}
Table~\ref{tab:weight_ratio} shows that under a fixed 100\,KB budget, lower update ratio enables more trainable layers and better accuracy; 0.2 (21 layers) gives the best result (88.78\%). Table~\ref{tab:memory_constraint} shows that with ratio 0.2, accuracy improves as memory increases and peaks at 90.27\% (320\,KB), then drops when early generic layers become trainable.

\subsubsection{Effectiveness of Dynamic Strategy}
Table~\ref{tab:dynamic_method} compares dynamic update with representative baselines on CIFAR-10. Although full fine-tuning reaches 91.60\%, it requires 2.1\,MB and is impractical for edge devices. Under the same 100\,KB budget, our dynamic method achieves 88.78\%, outperforming fixed channel update by 1.54\%.

\subsubsection{Evaluation on Downstream Tasks}
Table~\ref{tab:evaluation} compares with SparseUpdate~\cite{lin2022device} on eight datasets. Our FP8 dynamic method (Ours (F8D)) achieves 73.0\% average accuracy, improving over SparseUpdate (INT8) by 1.57\%, and consistently outperforms our FP8 fixed-update variant.

\section{Conclusion}
\label{sec:conclusion}

This work presents a sub-100\,KB on-device CNN training accelerator through hardware-software co-design. Mixed-level pruning compresses the base model by 95.2\%, and dynamic sparse update with activation pruning cuts activation memory to 3.4\% while fused group-normalization/convolution and row-wise sparse execution reduce training latency by 45\%. Implemented in TSMC 28\,nm at FP8 precision, the accelerator runs at 800\,MHz and achieves 8\,TOPS peak throughput, 31.13\,TOPS/W energy efficiency, and 14.8\,TOPS/mm$^2$ area efficiency, demonstrating practical edge training under tight resource constraints.

\bibliographystyle{IEEEtran}

\bibliography{references}

\end{document}